\documentclass[]{aa}
\usepackage{amsmath}
\usepackage{pifont}
\usepackage{graphicx}
\usepackage{txfonts}
\usepackage{color}
\usepackage[draft]{hyperref}
\usepackage{natbib}
\bibpunct{(}{)}{;}{a}{}{,} % to follow the A&A style
\usepackage[english]{babel}
\newcommand{\beqa}{\begin{eqnarray}} 
\newcommand{\cmark}{\ding{51}}%
\newcommand{\xmark}{\ding{55}}%
\newcommand{\eeqa}{\end{eqnarray}}

\newcommand{\bsub}{\begin{subequations}}
\newcommand{\esub}{\end{subequations}}
\newcommand{\beal}{\begin{align}}
\newcommand{\ealn}{\end{align}}

\usepackage{mathtools,tikz,caption}
\DeclareRobustCommand\sampleline[1]{%
  \tikz\draw[#1] (0,0) (0,\the\dimexpr\fontdimen22\textfont2\relax)
  -- (2em,\the\dimexpr\fontdimen22\textfont2\relax);%
}

\def\gsim{\mathrel{\rlap{\lower 4pt \hbox{\hskip 1pt $\sim$}}\raise 1pt \hbox {$>$}}}
\def\lsim{\mathrel{\rlap{\lower 4pt \hbox{\hskip 1pt $\sim$}}\raise 1pt \hbox {$<$}}}

\usepackage[separate-uncertainty=true]{siunitx}
\usepackage{graphicx}
\usepackage{txfonts}
\usepackage{todonotes}
\usepackage[flushleft]{threeparttable}


\begin{document} 

\titlerunning{Cobalt and iron in Type\,Ia supernovae}

\title{Limits on stable iron in Type\,Ia supernovae from NIR spectroscopy\thanks{Based on observations collected at the La Silla and Paranal
observatories of the European Southern Observatory, Chile in time allocated to proposals 290.D-5035 and 089.D-0647(A)}}

\authorrunning{A. Fl\"ors}
\author{\textbf{A. Fl\"ors\inst{1,2,3}
        \and J. Spyromilio\inst{1}
        \and K. Maguire\inst{4}
        \and S. Taubenberger\inst{1,2}
        \and W.~E. Kerzendorf\inst{1}
        \and S. Dhawan\inst{5}  
    }}

\institute{
    European Southern Observatory, Karl-Schwarzschild-Stra\ss e 2,
    D-85748 Garching bei M\"unchen, Germany \\
    \email{afloers@eso.org}
    \and Max-Planck-Institut f\"ur Astrophysik, Karl-Schwarzschild-Stra\ss e 1, D-85748 Garching bei M\"unchen, Germany
    \and Physik-Department, Technische Universit\"at M\"unchen, James-Franck-Stra\ss e 1, D-85748 Garching bei M\"unchen, Germany 
    \and Astrophysics Research Centre, School of Mathematics and Physics, Queen's University Belfast, Belfast BT7 1NN, UK
    \and Oskar Klein Centre, Department of Physics, Stockholm University,
    SE 106 91 Stockholm, Sweden 
} 

\date{Received 28 May 2018 / Accepted 10 October 2018}

\offprints{A. Fl\"ors}

\abstract{We obtained optical and near-infrared spectra of Type\,Ia supernovae (SNe\,Ia) at epochs ranging from 224 to 496 days after the explosion. The spectra show emission lines from forbidden transitions of singly ionised iron and cobalt atoms. We used non-local thermodynamic equilibrium (NLTE) modelling of the first and second ionisation stages of iron, nickel, and cobalt to fit the spectra using a sampling algorithm allowing us to probe a broad parameter space. We derive velocity shifts, line widths, and abundance ratios for iron and cobalt. The measured line widths and velocity shifts of the singly ionised ions suggest a shared emitting region. Our data are fully compatible with radioactive \element[][56]{Ni} decay as the origin for cobalt and iron. We compare the measured abundance ratios of iron and cobalt to theoretical predictions of various SN Ia explosion models. These models include, in addition to \element[][56]{Ni}, different amounts of \element[][57]{Ni} and stable \element[][54, 56]{Fe}. We can exclude models that produced only \element[][54, 56]{Fe} or only \element[][57]{Ni} in addition to \element[][56]{Ni}. If we consider a model that has \element[][56]{Ni}, \element[][57]{Ni}, and \element[][54, 56]{Fe} then our data imply that these ratios are \element[][54, 56]{Fe}\,/\,\element[][56]{Ni} $=0.272\pm0.086$ and \element[][57]{Ni}\,/\,\element[][56]{Ni} $=0.032\pm0.011$.}

%%-----------------

\maketitle

%
%________________________________________________________________

\section{Introduction}
Type\,Ia supernovae (SNe\,Ia) are a remarkably uniform class of objects. Exceptions such as overly bright \citep[e.g. SN\,1991T][]{1992ApJ...384L..15F,1992AJ....103.1632P,1992ApJ...387L..33R} or overly faint \citep[e.g. SN\,1991bg][]{1992AJ....104.1543F,1993AJ....105..301L,1996MNRAS.283....1T} supernovae have been extensively studied \citep[see][for a recent overview of the various Type\,Ia subtypes]{2017hsn..book..317T}. High-cadence all sky surveys \citep[e.g.][]{2012arXiv1202.2381K,2015MNRAS.451.4238S,2016arXiv161205560C,2017PASP..129j4502K} have discovered that members of the class exhibit a broad spectrum of disorders. However, the Branch-normal Type\,Ias \citep{2006PASP..118..560B} remain the dominant detected class and one of the best distance indicators at the disposal of astronomers. Following calibration procedures, for example the one by \citet{1993ApJ...413L.105P}, that correlate the width of the light curve with the peak brightness, SNe\,Ia exhibit a very small dispersion in their absolute magnitudes. As such, their use in cosmology is extensive and understanding them is of general interest \citep{2018SSRv..214...57L}.

\citet{1969ApJ...157..623C} proposed powering of the optical and infrared (IR) displays of SNe to originate in energy deposition from the decay of radioactive \element[][56]{Ni} to \element[][56]{Co} and subsequently to \element[][56]{Fe}. Depending on the temperature and density, burning to nuclear statistical equilibrium converts much of the progenitor white dwarf to iron group elements (\citealp{1960ApJ...132..565H}; see \citealp{2013MNRAS.429.1156S} for a recent calculation of yields). Many studies have provided both direct and indirect evidence for this scenario. Direct evidence can be found in \citet{2014Natur.512..406C} who detected $\gamma$-ray lines from the decay of \element[][56]{Co} at 847 and 1238\,keV. Indirect evidence was shown by \citet{1994ApJ...426L..89K} who, in Type\,Ia late-time optical spectra $\approx$\,200 days, observed that the doubly ionised emission lines of Co and Fe evolved according to the expected ratio that would result from the production of Fe as the daughter product of the radioactive decay of \element[][56]{Co}. The total energy of the explosion maps onto the mass of nucleosynthesised \element[][56]{Ni} and the luminosity at maximum light is directly linked to the mass of \element[][56]{Ni}  \citep{1982ApJ...253..785A}. The determination of the mass of $^{56}$Ni based on optical and IR photometry and spectroscopy has been the subject of many papers and almost as many models. In general, of order 0.4\,M$_{\odot}$ to 0.8\,M$_{\odot}$ of \element[][56]{Ni} is found to be produced in Branch-normal SNe\,Ia (\citealp{1982ApJ...253..785A}; \citealp[for a more recent study see][]{2015MNRAS.454.3816C}). 

The post maximum optical and near-infrared (NIR) spectra of SNe\,Ia exhibit a plethora of iron group emission lines, predominantly in the singly and doubly ionised states \citep{1980PhDT.........1A,1989ApJ...343..323F,2005A&A...437..983K,2015ApJ...814L...2F}. Many authors have provided an extensive analysis of the physical conditions in the ejecta and the processes that generate the spectrum. Post-maximum spectra have been published by many authors and show remarkable similarity in features and evolution \citep{1998ApJ...499L..49M}. \citet{2010ApJ...708.1703M} made the startling discovery that often the singly ionised emission lines of the iron group elements exhibit a different systemic velocity when compared to their doubly ionised counterparts and proposed that there exists a region within the core of the ejecta where electron capture processes dominate the nucleosynthetic yield. 

In this paper, we fit late-time NIR observations with emission lines from NLTE level populations to derive the evolution of the mass ratio of \ion{Co}{II} to \ion{Fe}{II}. We aim to place limits on the stable Fe and \element[][57]{Ni} in the ejecta by observing the evolution in the mass ratio of Co to Fe. In Section \ref{SectionObservations}, we briefly describe the observations and in Section \ref{SectionFits}, we present our NLTE emission line fit code. In Section \ref{SectionFitResults}, we present the results of our spectral fits and derive the mass ratio $\mathrm{M_{\ion{Co}{II}}\,/\,M_{\ion{Fe}{II}}}$. In Section \ref{SectionBayes}, we compare the mass ratio results of our SN sample with explosion model predictions within the Bayesian framework. The results are discussed in Section \ref{SectionDiscussion}.

\section{Observations}
\label{SectionObservations}
\begin{figure*}
    \centering
    \begin{minipage}{.475\textwidth}
        \centering
        \includegraphics[width=1\linewidth]{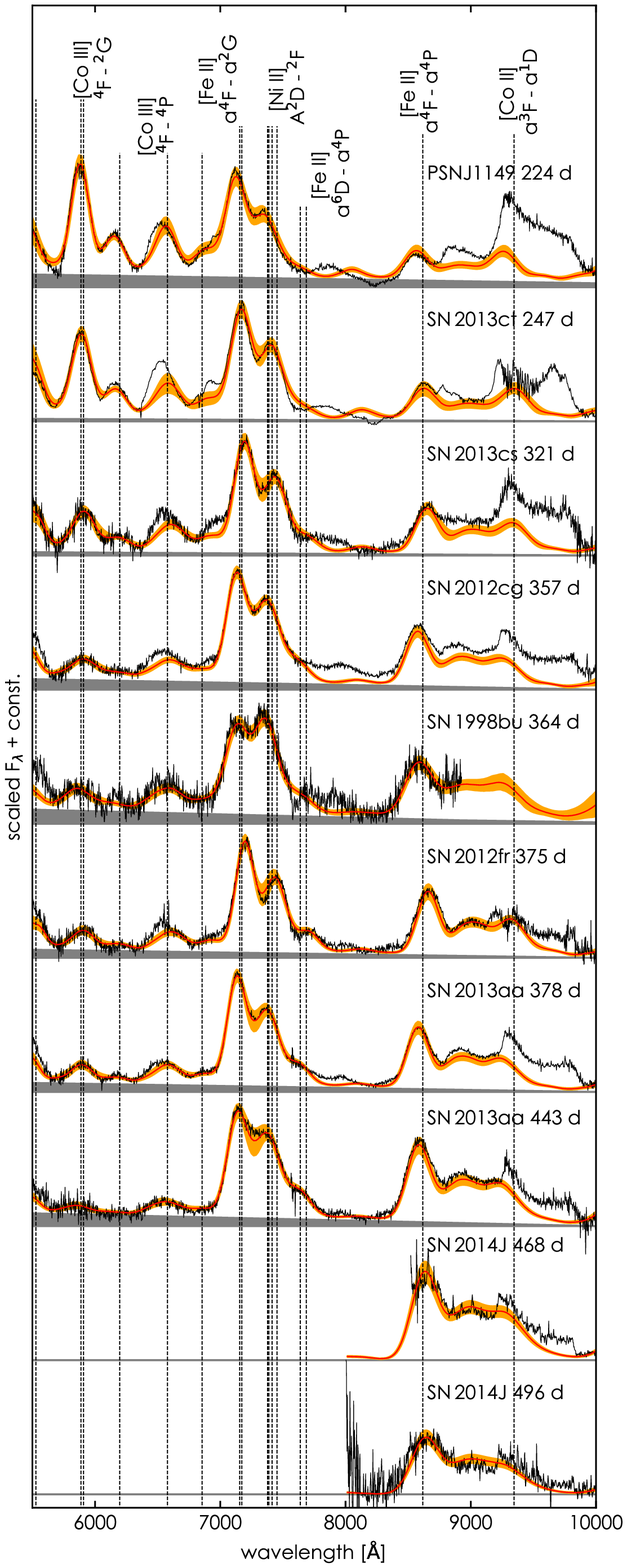}
    \end{minipage}%
    \begin{minipage}{.475\textwidth}
        \centering
        \includegraphics[width=1\linewidth]{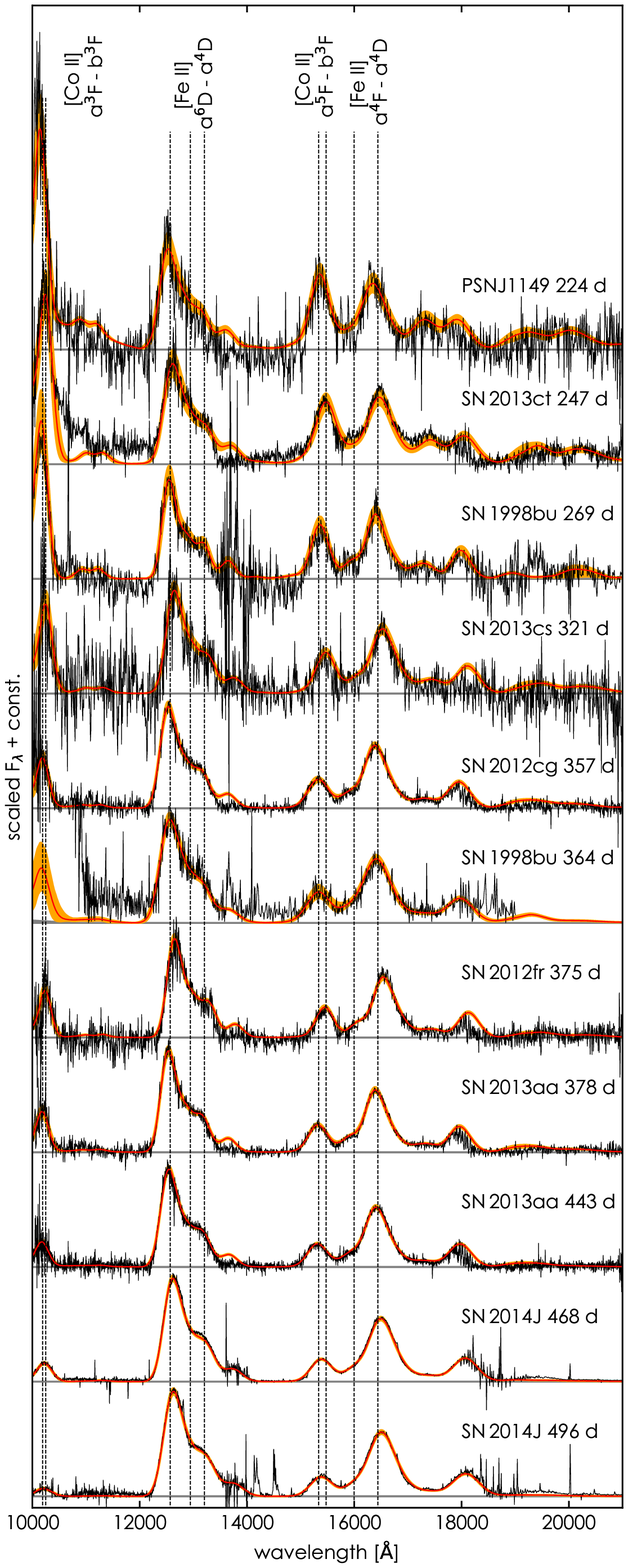}
    \end{minipage}
    \caption{Optical (left) and NIR (right) spectra of SNe\,Ia obtained with X-Shooter, FORS1, and ISAAC at the VLT, SOFI at the NTT, and GNIRS at Gemini-North. The spectra are arranged in epoch starting with the youngest at the top and corrected for redshift and extinction. Fluxes are normalized to the $7\,300\,$\AA\,[\ion{Fe}{II}]\,+\,[\ion{Ni}{II}] feature (optical) and $12\,600\,$\AA\,[\ion{Fe}{II}] feature (NIR). The red line indicates the mean flux of all fit models at each wavelength, the orange shaded area marks the $68\,\%$ uncertainty of the fit. Dashed vertical lines indicate the strongest lines as given in Table \ref{tableLines} ([\ion{Fe}{II}]: \sampleline{dashed}, [\ion{Ni}{II}]: \sampleline{dotted}, [\ion{Co}{II}]:  \sampleline{dash pattern=on .7em off .2em on .05em off .2em}, [\ion{Co}{III}]:  \sampleline{dash pattern=on .5em off .2em on .05em off .2em on .05em off .2em}). The subtracted background is shown as a grey band. Fits were performed for the optical and NIR spectra at the same time. The lines/ions composing the features below $5\,500$\,\AA\,are not included in our fits.}
    \label{OpticalSpectra}
\end{figure*}
Four sets of data are included in this work (see Table \ref{tableExtinctionRedshift}). One set from the GNIRS instrument on Gemini-North, two from the X-Shooter instrument  at ESO's Paranal observatory and the fourth from SOFI at the New Technology Telescope (NTT) and ISAAC+FORS1 at the Very Large Telescope (VLT). GNIRS, SOFI, and ISAAC cover the NIR bands, FORS1 covers the optical while X-Shooter covers both the optical and NIR bands. As we are mainly interested in the abundances of various radioactive isotopes all epochs in this work are given in days after the explosion, assuming a rise time of $\sim 18$ days \citep{2011MNRAS.416.2607G}.

We use two NIR spectra of SN2014J obtained with GNIRS at Gemini-North \citep[see][accepted]{2018arXiv180502420D}. Optical+NIR spectra of SN\,2013cg, SN\,2012fr, SN\,2013aa, SN\,2013cs and SN\,2013ct were obtained by \citet{2016MNRAS.457.3254M} at the VLT. \citet{2018MNRAS.477.3567M} also obtained a spectrum of PSNJ11492548-0507138 (hereafter PSNJ1149) and a second spectrum of SN\,2013aa. We also use two spectra of SN\,1998bu \citep{2004A&A...426..547S}. For details of the reduction see the publication papers.

A number of broad ($\approx7\,000$ to $9\,000$ km\,s$^{-1}$) emission lines at 1.55\,$\mu$m and 1.65\,$\mu$m are evident in the data. We identify these as emission by the 1.644\,$\mu$m [\ion{Fe}{II}] line arising from the a$^4$F--a$^4$D multiplet and the 1.547\,$\mu$m [\ion{Co}{II}] line from the a$^5$F--b$^3$F multiplet. The a$^4$F-a$^4$D [\ion{Fe}{II}] multiplet includes the 1.533\,$\mu$m [\ion{Fe}{II}] line. The identification of the various features has been extensively covered in the literature. The detection of the 1.257\,$\mu$m a$^6$D$_{9/2}$--a$^4$D$_{7/2}$ confirms the identification and strength of the a$^4$F$_{9/2}$--a$^4$D$_{7/2}$ 1.644\,$\mu$m line as they arise from the same upper level. Similarly the detection in many of our spectra of the 1.090\,$\mu$m (a$^3$F$_4$--b$^3$F$_4$) line of [\ion{Co}{II}] which shares an upper level with the 1.547\,$\mu$m (a$^5$F$_4$--b$^3$F$_4$) line also secures the identification.

\begin{table*} 
    \caption{Overview of spectra in our sample.}
    \label{tableExtinctionRedshift}     
    \centering 
    \scalebox{0.95}{
    \begin{tabular}{l c c cccccc} 
        \hline
        \hline 
        \centering 
        Supernova & E$(B-V)$\tablefootmark{a} & z\tablefootmark{b} & Date of max.& Epoch\tablefootmark{c} & Telescope & Instrument & Resolution & Source\\
        & (mag) & & & & & & $ \lambda / \Delta\lambda$\\
        \hline  
        SN 1998bu & 0.30\tablefootmark{d} & 0.002992 & 1998 May 21 & $269$\,d & NTT          & SOFI   & 500     & ~\,1\tablefootmark{e} \\
        &                       &          &          & $364$\,d & VLT          & ISAAC, FORS1 & 1500, 440 & ~\,1\tablefootmark{e} \\
        SN 2012cg & 0.20\tablefootmark{f} & 0.001458 & 2012 June 03 & $357$\,d & VLT          & X-Shooter & 6200/8800/5300\tablefootmark{g}   & 2                 \\
        SN 2012fr & 0.018                 & 0.005457 & 2012 Nov 12 & $375$\,d & VLT          & X-Shooter  & 6200/8800/5300\tablefootmark{g}  & 2                 \\
        SN 2013aa & 0.169                 & 0.003999 & 2013 Feb 21  & $378$\,d & VLT          & X-Shooter  & 6200/8800/5300\tablefootmark{g}  & 2                 \\
        &                       &          &          & $443$\,d & VLT          & X-Shooter & 6200/8800/5300\tablefootmark{g}   & 3                 \\
        SN 2013cs & 0.082                 & 0.009243 & 2013 May 26  & $321$\,d & VLT          & X-Shooter & 6200/8800/5300\tablefootmark{g}   & 2                 \\
        SN 2013ct & 0.025                 & 0.003843 & 2013 May 04  & $247$\,d & VLT          & X-Shooter & 6200/8800/5300\tablefootmark{g}   & 2                 \\ 
        SN 2014J  & 1.37\tablefootmark{h} & 0.000677 & 2014 Feb 01 & $468$\,d & Gemini-North & GNIRS  & 1800      & 4                 \\
        &                       &          &          & $496$\,d & Gemini-North & GNIRS   & 1800     & 4                 \\
        PSN J1149 & 0.025                 & 0.005589 & 2015 July 12 & $224$\,d & VLT          & X-Shooter & 6200/8800/5300\tablefootmark{e}   & 3                 \\
        \hline
    \end{tabular} }
    \tablebib{
        (1)~\citet{2004A&A...426..547S}; (2)~\citet{2016MNRAS.457.3254M}; (3)~\citet{2018MNRAS.477.3567M}; (4)~\citet{2018arXiv180502420D}.
    }
    \tablefoot{
        \tablefoottext{a}{MW E$(B-V)$ in magnitudes from \citet{2011ApJ...737..103S}. If additional host galaxy extinction is present we quote the combined Galactic and host galaxy E$(B-V)$ in magnitudes.}
        \tablefoottext{b}{Heliocentric redshifts are taken from the Nasa Extragalactic Database (NED).}
        \tablefoottext{c}{Days after the explosion, assuming a rise time of $\sim 18$ days \citep{2011MNRAS.416.2607G}. The spectra were scaled to match the photometry at this phase.}
        \tablefoottext{d}{\citet{1999ApJS..125...73J} report the extinction towards the SN as $A_V=0.94\,$mag.}
        \tablefoottext{e}{We subtracted a light echo as found by \citet{2001ApJ...549L.215C} scaled up by a factor of two \citep[see][]{2004A&A...426..547S}.}      
        \tablefoottext{f}{\citet{2012ApJ...756L...7S} found host galaxy extinction of E$(B-V)=0.18\,$mag.}
        \tablefoottext{g}{Resolution of the three X-Shooter arms UVB/VIS/NIR}
        \tablefoottext{h}{\citet{2014ApJ...788L..21A} determine E$(B-V)=1.37\,$mag with low total-to-selective extinction R$_V=1.4$.}
    }
\end{table*}
\section{Fitting the spectra}
\label{SectionFits}
\begin{figure}
        \centering
    \resizebox{\hsize}{!}{\includegraphics{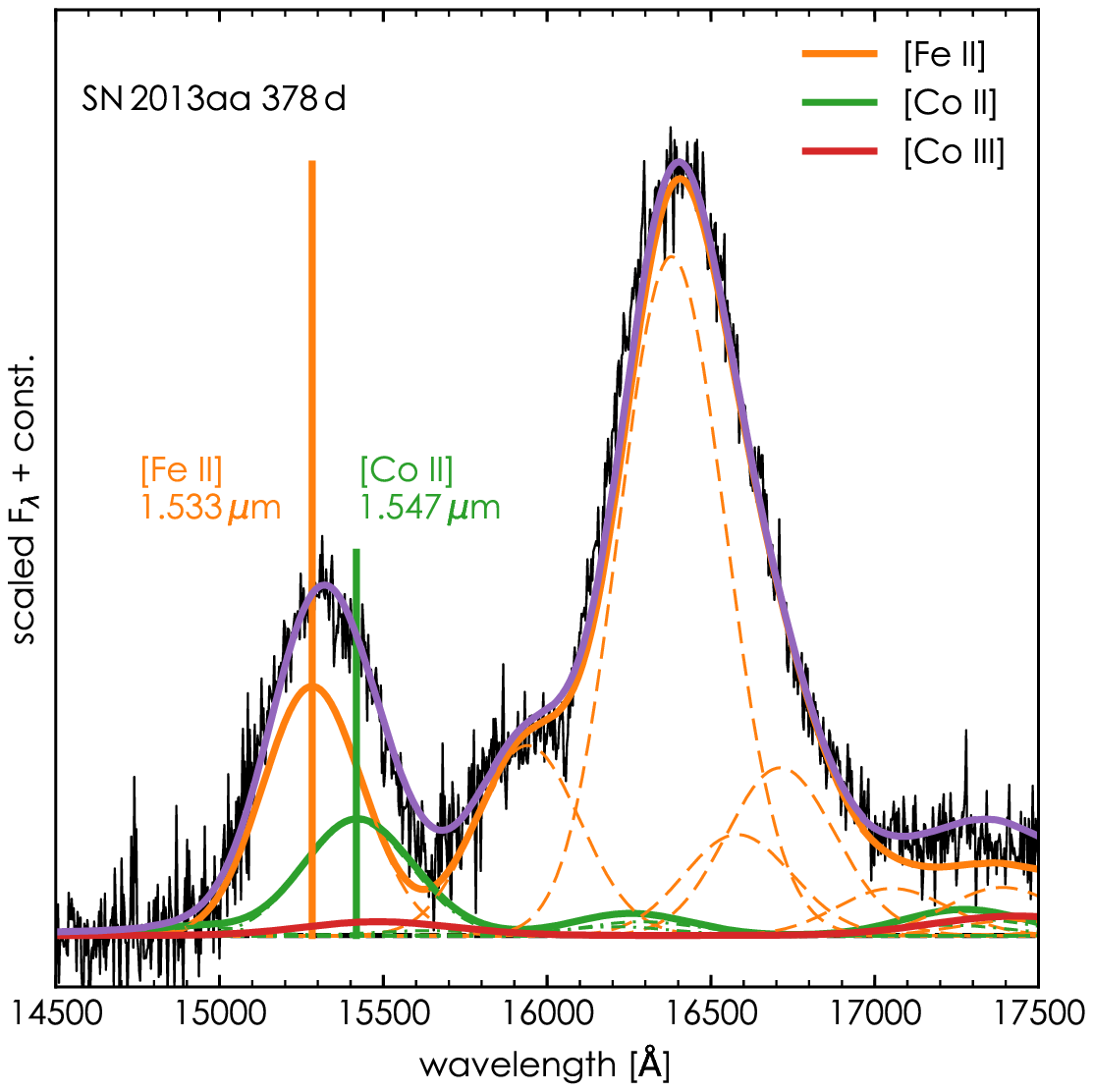}}
        \caption{
            Best fit model of SN\,2013aa at 378\,d in the H-band. Shown is the extinction and redshift corrected spectrum (black line) and individual contributions from the ions \ion{Fe}{II} (orange), \ion{Co}{II} (green) and \ion{Co}{III} (red) as solid curves. The purple line is the combined emission of the three ions. The two solid vertical lines are used to compute $\mathrm{M_{\ion{Co}{II}}\,/\,M_{\ion{Fe}{II}}}$. Dashed curves indicate the contribution of the individual lines in the blended region. We do not show \ion{Ni}{II} as it does not have any lines in this region.}
        \label{2014Jexample}
\end{figure}
\subsection{Emission line model}
We use a one zone model with \ion{Fe}{II}, \ion{Ni}{II}, and \ion{Co}{II}; it also includes \ion{Co}{III} if the spectrum covers the $6\,000$\,\AA\,region. For this set of ions, we solve the NLTE rate equations treating only collisional excitation and de-excitation by collisional and radiative processes to compute energy level populations and derive forbidden line emissivities. For the collisional processes we use a thermal electron gas characterised by a Boltzmann distribution and an electron density. We ignore radiative transfer effects as the optical depths of the lines under consideration are very low at the observed epochs. We also do not consider non-thermal excitations as the energy going into this channel at the relatively high electron densities we determine is also very low \citep{1989ApJ...343..323F}. We do not include charge exchange and time-dependent terms in the NLTE rate equations. Even though we do not treat continuum processes directly, we include a parametrized smooth continuum in the model (see section \ref{SecParameterEstimation}). We use such a smooth continuum to include uncertainties due to true continuum processes in the early spectra and potentially low-level light echoes. We do not treat energy deposition in the ejecta, and as a result our model does not match the emitted radiation over all wavelengths, but only for certain lines. We do not solve the ionisation balance but treat the number of emitting ions as a fit parameter. Effectively, we solve a simplified and parametrized NLTE problem. 

A one-zone model, by default spherically symmetric, convolved with a Gaussian line profile serves to fit the spectra and extract the salient properties of the lines (flux, rest velocity, and Doppler broadening). One evident limitation of the one-zone model is that different ionisation stages, not necessarily co-located in the ejecta (see section \ref{sectionShiftsWidths}), are likely to have different excitation conditions in some fraction of the emitting regions. This work, which concentrates on the singly ionised species, is not significantly affected by this limitation.

The features at $4700$ and $5200\,$\AA\,are a blend of singly and doubly ionised  iron, with possibly asymmetric line profiles. Changes of the velocity offset and width of the $4\,700$\,\AA\, feature with time suggest that this wavelength region is not optically thin until $400$ days after the explosion \citep{2016MNRAS.462..649B}. A detailed analysis of this feature would require a more sophisticated model process. As [\ion{Fe}{III}] only slightly influences the rest of the Optical+NIR spectrum we do not attempt to fit spectral features below $5\,500$\,\AA.

We fit the $5\,900$\,\AA\,region with [\ion{Co}{III}] and the double-peaked feature between $7\,000$ and $7\,800$\,\AA\,with [\ion{Fe}{II}]+[\ion{Ni}{II}]. All spectra exhibit a strong emission feature between $8\,500\,$\AA\,and $9\,900\,$\AA. We can reproduce the feature at $8\,600\,$\AA\,with lines from the \ion{Fe}{II} a$^4$F--a$^4$P and \ion{Co}{II} a$^3$F--b$^3$F multiplets, but in spectra which are less than one year old the red part of the feature between $8\,800\,$\AA\,and $9\,900\,$\AA\,cannot be well explained by emission from only ions in our sample. Models by \citet{2017ApJ...845..176B}, among other authors, suggest the presence of \ion{S}{III} in this region. We only use the $8\,600\,$\AA\, feature for spectra of SN\,2014J, but we do not use the $8\,800$ to $9\,900\,$\AA\,region. 

The strongest lines of the considered ions are given in Table \ref{tableLines}. An overview of the model ions with their respective atomic data is given in Table \ref{tableAtomicData}. For \ion{Fe}{II} we use the atomic data from \citet{2015ApJ...808..174B} as they provide collision strengths for the higher levels responsible for optical transitions. The spectra are displayed in Fig.~\ref{OpticalSpectra} in chronological order. The data are corrected for the redshift of the host, as well as Galactic, and, if applicable, host galaxy extinction according to Table \ref{tableExtinctionRedshift}. The simple model of \ion{Fe}{II}, \ion{Co}{II}, \ion{Co}{III} and \ion{Ni}{II} fits the data well above $5\,500$\,\AA.  
\begin{table} 
        \caption{Strongest lines of the included ions in the optical and NIR.} 
        \label{tableLines}      
        \centering 
        \begin{tabular}{l c  c } 
        \hline
        \hline 
        $\lambda_{\text{rest}}$($\mu$m) & Ion & Transition\\
                \hline 
        0.5528 & [\ion{Fe}{II}] & a$^4$F$_{7/2}\,-\,$a$^2$D$_{5/2}$\\
        0.5888 & [\ion{Co}{III}] & a$^4$F$_{9/2}\,-\,$a$^2$G$_{9/2}$\\
        0.5908 & [\ion{Co}{III}] & a$^4$F$_{7/2}\,-\,$a$^2$G$_{7/2}$\\
        0.6197 & [\ion{Co}{III}] & a$^4$F$_{7/2}\,-\,$a$^2$G$_{9/2}$\\
        0.6578 & [\ion{Co}{III}] & a$^4$F$_{9/2}\,-\,$a$^4$P$_{5/2}$\\
        0.6855 & [\ion{Co}{III}] & a$^4$F$_{7/2}\,-\,$a$^4$P$_{3/2}$\\
        0.7155 & [\ion{Fe}{II}] & a$^4$F$_{9/2}\,-\,$a$^2$G$_{9/2}$\\
        0.7172 & [\ion{Fe}{II}] & a$^4$F$_{7/2}\,-\,$a$^2$G$_{7/2}$\\
        0.7378 & [\ion{Ni}{II}] & z$^2$D$_{5/2}\,-\,$a$^2$F$_{7/2}$\\
                0.7388 & [\ion{Fe}{II}] & a$^4$F$_{5/2}\,-\,$a$^2$G$_{7/2}$\\
        0.7414 & [\ion{Ni}{II}] & z$^2$D$_{3/2}\,-\,$a$^2$F$_{5/2}$\\
        0.7453 & [\ion{Fe}{II}] & a$^4$F$_{7/2}\,-\,$a$^2$G$_{9/2}$\\
        0.7638 & [\ion{Fe}{II}] & a$^6$D$_{7/2}\,-\,$a$^4$P$_{5/2}$\\
        0.7687 & [\ion{Fe}{II}] & a$^6$D$_{5/2}\,-\,$a$^4$P$_{3/2}$\\
        0.8617 & [\ion{Fe}{II}] & a$^4$F$_{9/2}\,-\,$a$^4$P$_{5/2}$\\
        0.9345 & [\ion{Co}{II}] & a$^3$F$_{3}\,-\,$a$^1$D$_{2}$\\
        1.0190 & [\ion{Co}{II}] & a$^3$F$_{4}\,-\,$b$^3$F$_{4}$\\
        1.0248 & [\ion{Co}{II}] & a$^3$F$_{3}\,-\,$b$^3$F$_{3}$\\
        1.2570 & [\ion{Fe}{II}] & a$^6$D$_{9/2}\,-\,$a$^4$D$_{7/2}$\\
        1.2943 & [\ion{Fe}{II}] & a$^6$D$_{5/2}\,-\,$a$^4$D$_{5/2}$\\
        1.3206 & [\ion{Fe}{II}] & a$^6$D$_{7/2}\,-\,$a$^4$D$_{7/2}$\\
        1.5335 & [\ion{Fe}{II}] & a$^4$F$_{9/2}\,-\,$a$^4$D$_{5/2}$\\
        1.5474 & [\ion{Co}{II}] & a$^5$F$_{5}\,-\,$b$^3$F$_{4}$\\
        1.5488 & [\ion{Co}{III}] & a$^2$G$_{9/2}\,-\,$a$^2$H$_{9/2}$\\
        1.5995 & [\ion{Fe}{II}] & a$^4$F$_{7/2}\,-\,$a$^4$D$_{3/2}$\\
        1.6440 & [\ion{Fe}{II}] & a$^4$F$_{9/2}\,-\,$a$^4$D$_{7/2}$\\
                \hline
        \end{tabular}
\end{table}
\begin{table} 
        \caption{Ions included in the fits and their atomic data sets.} 
        \label{tableAtomicData} 
        \centering 
        \footnotesize
        \begin{tabular}{c c  c c} 
        \hline
        \hline 
        Ion                     & Levels\tablefootmark{a} & Ref. $A_{ij}$\tablefootmark{b}   & Ref. $\Upsilon_{ij}$\tablefootmark{c} \\    
                \hline  
                $\ion{Fe}{II}$  & 52                                      & 1 & 1           \\
                $\ion{Co}{II}$  & 15                                      & 2 & 2           \\
                $\ion{Co}{III}$ & 15                                      & 3 & 3           \\
                $\ion{Ni}{II}$  & 18                                      & 4 & 5           \\
                \hline
        \end{tabular}
    \tablebib{
        (1)~\citet{2015ApJ...808..174B}; (2)~\citet{2016MNRAS.456.1974S}; (3)~\citet{2016MNRAS.459.2558S}; (4)~\citet{2016A&A...587A.107C}; (5)~\citet{2010AA...513A..55C}
        }
        \tablefoot{
                \tablefoottext{a}{Energy levels and statistical weights are taken from NIST \citep{NIST_ASD}.}
                \tablefoottext{b}{Einstein $A$ coefficient between levels $i$ and $j$.}
                \tablefoottext{c}{Maxwellian averaged collisional strength between levels $i$ and $j$.}
        }
\end{table}
\subsection{Parameter estimation}
\label{SecParameterEstimation}
We explore the parameter space using the nested-sampling algorithm Nestle \citep[https://github.com/kbarbary/nestle, see also][]{2007MNRAS.378.1365S}. The algorithm allows us to sample from our flat priors over a large range by optimising the selection of variables based on the quality of the earlier fits. To compare the model with the data we assume a $\chi^2$ likelihood. Our set of four ions shares the same temperature and electron density. Each ion is allowed to have its own line width, velocity offset, and strength. Uniform priors are used for all parameters except the electron density, for which we use a log-uniform distribution.
There exists a significant body of work in the literature starting with \citet{1980PhDT.........1A}, and a plethora of other authors \citep[see e.g.][]{1994ApJ...426L..89K,2005A&A...437..983K,2015ApJ...814L...2F,2017ApJ...845..176B} that have shown that  temperatures for the ejecta of a SNe Ia during the first year after explosion lie in the range of $2\,000-15\,000$\,K. For our modelling we adopt the following boundaries for the priors: temperature range ($2\,000$ to $15\,000$\,K), electron densities between $10^{4}$ to $10^{7}$\,cm$^{-3}$, Doppler widths between $2$ and $15\times 10^{3}$\,km\,s$^{-1}$, and shifts between $-3$ and $3\times 10^{3}$\,km\,s$^{-1}$. The very large allowed range for the temperature prior does not affect the resulting fit. We note the presence of a weak continuum in some spectra bluewards of $9500\,$\AA; in those cases, we subtract a linear background (grey bands in Fig. \ref{OpticalSpectra}) in the optical. Widths and shifts of individual ions are determined using a global fit where the individual species are allowed to vary within the prior space.

Besides the fit parameters noted above, the fits provide us with the individual emission line emissivities. The ratios of the transitions arising from within the a$^4$F--a$^4$D \ion{Fe}{II} multiplet show little variation with temperature as the difference in the energy levels within a$^4$D is less than 1\,000\,cm$^{-1}$ and for the strongest lines at 1.644\,$\mu$m and 1.533\,$\mu$m less than 500\,cm$^{-1}$. The higher-lying states responsible for lines in the optical vary significantly in strength over the considered temperature range and thus allow us to estimate the electron temperature. However, there is a degeneracy between electron density and temperature. Good fits can be achieved with high temperature and low densities, and vice versa. Some of these high temperatures and low densities would not be compatible with a prior of $\sim 0.4 -0.8\,$M$_\odot$ of iron group elements. We have not adopted such a prior for the fitting as absolute mass estimates suffer considerably from systematics (e.g. distance). We want to emphasise that the uncertainties from this degeneracy are much smaller than the priors (see Table \ref{tableFitResults}).

\section{Fit results}
\label{SectionFitResults}
\subsection{Fit parameters and $\mathrm{M_{\ion{Co}{II}}\,/\,M_{\ion{Fe}{II}}}$}

The ratio of the $H$-band [\ion{Co}{II}] to [\ion{Fe}{II}] lines places limits on the temperature of the ejecta of above $1\,500$\,K. Below this temperature, the blue wing of the 1.55\,$\mu$m spectral feature is underproduced and the blend of the 1.634 [\ion{Co}{II}] and 1.644 [\ion{Fe}{II}] lines which accounts for the 1.65-$\mu$m feature is unable to reproduce the spectra. The presence of the shoulder from the a$^4$F$_{7/2}-$a$^4$D$_{3/2}$ [\ion{Fe}{II}] line at 1.599\,$\mu$m in spectra at all epochs indicates that the emission originates from a high-density region ($n_e > 10^5$\,cm$^{-3}$).

The absolute mass of the emitting material strongly depends on the assumed temperature in the ejecta and the distance to the object. We do not consider that our model is sophisticated enough to provide detailed constraints on these parameters (see for example \citet{2015ApJ...814L...2F} or \citet{2018ApJ...861..119D} for such work). On the other hand, mass ratios are only weakly dependent on the temperature. In particular, from the line ratio of 1.547\,$\mu$m [\ion{Co}{II}] to 1.533\,$\mu$m  [\ion{Fe}{II}] (see Fig. \ref{2014Jexample}) we can determine the mass fraction of \ion{Co}{II} to \ion{Fe}{II} \citep{1990MNRAS.245..570V}:

\begin{equation} 
        \frac{M_{\ion{Co}{II}}}{M_{\ion{Fe}{II}}}{\left(t\right)}  = \frac{F_{1.547}}{F_{1.533}}{\left(t\right)} \times \frac{1.533\,A_{1.533}\,g_{\ion{Fe}{II}}\,Z_{\ion{Co}{II}}{\left(t\right)}\,m_{\ion{Co}{II}}}{1.547\,A_{1.547}\,g_{\ion{Co}{II}}\,Z_{\ion{Fe}{II}}{\left(t\right)}\,m_{\ion{Fe}{II}}}\times e^{\frac{k_\text{B}\,2\,040\,\text{K}}{k_\text{B}\,T{\left(t\right)}}}
        \label{coIIfeIIratio}
,\end{equation}
where $T$ is the temperature of the gas in degrees Kelvin reflecting the difference in the upper energy levels of the two transitions, $A$ are the transition probabilities between the a$^4$F$_{9/2}-$a$^4$D$_{5/2}$ \ion{Fe}{II} and a$^5$F$_5-$b$^3$F$_4$ \ion{Co}{II} levels, $g$ are the statistical weights of the upper levels, $m$ are the atomic masses and
\begin{equation} 
        Z{\left(t\right)} = \sum_i \left( g_i\,\exp{[-E_i/ (k_{\text{B}}T{\left(t\right)})]}\right)
        \label{PartitionFunction}
\end{equation}
are the atomic partition functions. The flux ratio, partition functions, temperature and the inferred mass ratio are functions of time after explosion. For the atomic data used in this work Eq.~\ref{coIIfeIIratio} reduces to:
\begin{equation} 
        \frac{M_{\ion{Co}{II}}}{M_{\ion{Fe}{II}}}{\left(t\right)}  = 0.0613 \times \frac{F_{1.547}}{F_{1.533}}{\left(t\right)} \times \frac{Z_{\ion{Co}{II}}}{Z_{\ion{Fe}{II}}}{\left(t\right)}\times e^{\frac{k_\text{B}\,2\,040\,\text{K}}{k_\text{B}\,T{\left(t\right)}}}
        \label{CoIIFeIIreduced}
.\end{equation}

\begin{figure*}
        \centering
        \includegraphics[width=18cm]{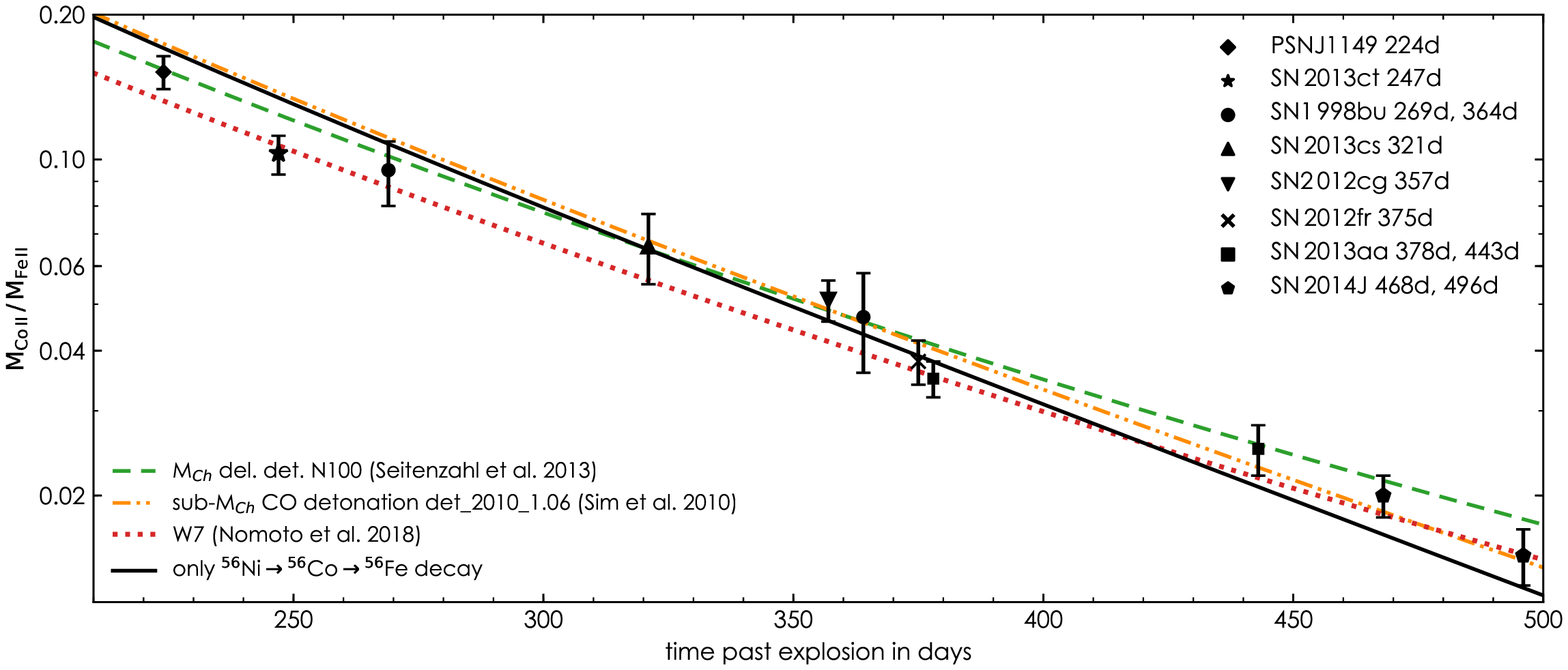}
        \caption{
            Evolution of the inferred $\mathrm{M_{\ion{Co}{II}}\,/\,M_{\ion{Fe}{II}}}$ ratio with time. We assumed a rise time of 18 days \citep{2011MNRAS.416.2607G}. The error bars reflect the $68\%$ posterior interval of the mass ratio. The coloured lines show the expected mass ratio $\mathrm{M_{Co}\,/\,M_{Fe}}$ of the M$_{Ch}$ delayed-detonation model `N100' \citep[][green]{2013MNRAS.429.1156S}, the sub-M$_{Ch}$ CO detonation model `det\_2010\_1.06' \citep[][orange]{2010ApJ...714L..52S} and the M$_{Ch}$ `W7 Z$_\odot$' model \citep[][red]{2018SSRv..214...67N}. The black line is not a fit to the data and represents the $\mathrm{M_{Co}\,/\,M_{Fe}}$ ratio assuming only radioactive decay from \element[][56]{Ni} to \element[][56]{Co} to \element[][56]{Fe}. The same ionisation fractions of \ion{Co}{II} and \ion{Fe}{II} allow us to generalise $\mathrm{M_{\ion{Co}{II}}\,/\,M_{\ion{Fe}{II}}}$ to $\mathrm{M_{Co}\,/\,M_{Fe}}$ and compare the two ratios (see Section \ref{SectionBayes}).} 
        \label{decay}
\end{figure*}
As the SN\,1998bu 269\,d spectrum covers only the NIR we cannot determine the temperature and electron density from the ratio of optical to NIR \ion{Fe}{II} lines. To determine the line widths and velocity offsets we fit the spectrum for a fixed temperature $T=7\,200\,$K and electron density log\,(n$_{e}$\,/\,cm$^{-3})=6.2$. These numbers lie between the fit results of SN\,2013ct at 247\,d and SN\,2013cs at 321\,d. The mass ratio is then computed for $T=(7200\pm1500)\,$K and log\,(n$_{e}$\,/\,cm$^{-3})=(6.20\pm0.15)$. The uncertainties of the density and temperature are reflected in an increased uncertainty of the derived mass ratio M$_{\ion{Co}{II}}$\,/\,M$_{\ion{Fe}{II}}$ for this spectrum.

The spectrum of SN1998bu at 364\,d does not cover the $1.019$\,$\mu$m \ion{Co}{II} line, which affects the uncertainty of the inferred mass ratio significantly as the \ion{Co}{II} 1.547\,$\mu$m line is rather weak compared to the \ion{Fe}{II} 1.533\,$\mu$m line. Good fits can be achieved for a wide range of cobalt masses in this case.  

The spectra of SN\,2014J only cover the NIR wavelengths down to $8\,500\,$\AA. We therefore cannot use the $7\,200\,$\AA\,[\ion{Fe}{II}]\,+\,[\ion{Ni}{II}] complex to determine the density and temperature in the ejecta. We find that the $8\,600\,$\AA\,feature is well reproduced (best at late epochs $>350$\,d) by our emission model for other SNe in our sample with both optical and NIR wavelength coverage. We estimate the temperature and density of our SN\,2014J spectra by fitting the $8\,600\,$\AA\,feature. 

In Table \ref{tableFitResults}, we present the $68\%$ credibility interval for the electron density, temperature, \ion{Co}{II}\,/\,\ion{Fe}{II} mass ratio, widths and shifts of [\ion{Fe}{II}], [\ion{Co}{II}] and [\ion{Co}{III}]. The evolution of the mass ratio with the time since the supernova explosion is shown in Fig.~\ref{decay}. The error bars in the plot reflect mainly the uncertainty of the degenerate density and temperature, and, for very late spectra ($>400$\,d), the Doppler width of singly ionised cobalt. Superimposed on the  data points is the change of the mass ratio of cobalt to iron as expected from the radioactive decay of \element[][56]{Ni} (black curve). The line is not a fit to the data but rather the evolution of the mass ratio assuming the established decay half-lives of 6.08 and 77.27 days for \element[][56]{Ni} and \element[][56]{Co} respectively: 
\begin{align}
        &_{28}^{56}\text{Ni} \xrightarrow{\,\,\,\,T_{1/2}=6.08\,\text{d}\,\,\,\,}\,_{27}^{56}\text{Co} \xrightarrow{\,\,\,\,\,\,T_{1/2}=77.27\,\text{d}\,\,\,\,\,}\, _{26}^{56}\text{Fe.}\label{Ni56Decay}
\end{align}

\begin{table*} 
        \tiny
        \caption{$68\%$ posterior probability intervals of the fit parameters.} 
        \label{tableFitResults} 
        \centering 
    \scalebox{.9}{
        \begin{tabular}{l c  c c c c c c c c c } 
                \hline
                \hline 
                SN        & Age\tablefootmark{a}  &  Mass Ratio & log [n$_{e}$ / cm$^{-3}$] & T & Width [\ion{Fe}{II}]\tablefootmark{b} & Shift [\ion{Fe}{II}]\tablefootmark{c} & Width [\ion{Co}{II}]\tablefootmark{b} & Shift [\ion{Co}{II}]\tablefootmark{c} & Width [\ion{Co}{III}]\tablefootmark{b} & Shift [\ion{Co}{III}]\tablefootmark{c}\\     
                          & [days]  &  M$_{\ion{Co}{II}}$\,/\,M$_{\ion{Fe}{II}}$  &  & [$10^3$\,K] & [$10^3$\,km\,s$^{-1}$] & [$10^3$\,km\,s$^{-1}$] & [$10^3$\,km\,s$^{-1}$] & [$10^3$\,km\,s$^{-1}$] & [$10^3$\,km\,s$^{-1}$] & [$10^3$\,km\,s$^{-1}$]\\
                \hline  
                PSN J1149 & 224 & $0.152^{+0.012}_{-0.012}$ & $6.39^{+0.10}_{-0.08}$ & $8.5^{+2.3}_{-1.3}$ & $7.88^{+0.65}_{-0.55}$ & $-1.74^{+0.26}_{-0.26}$ & $7.21^{+0.99}_{-0.77}$ & $-2.10^{+0.36}_{-0.37}$ & $8.63^{+0.29}_{-0.28}$  & $-0.71^{+0.17}_{-0.16}$\\
                SN 2013ct & 247 & $0.103^{+0.010}_{-0.009}$ & $6.42^{+0.28}_{-0.15}$ & $7.6^{+1.6}_{-1.6}$ & $8.19^{+0.53}_{-0.47}$ & $0.28^{+0.20}_{-0.22}$  & $7.78^{+0.76}_{-0.80}$ & $0.78^{+0.43}_{-0.39}$  & $9.31^{+0.34}_{-0.32}$  & $-0.14^{+0.18}_{-0.18}$\\
                SN 1998bu & 269 & $0.095^{+0.015}_{-0.014}$ & $6.20^{+0.15}_{-0.15}$ & $7.2^{+1.5}_{-1.5}$ & $6.51^{+0.44}_{-0.44}$ & $-0.94^{+0.32}_{-0.29}$ & $7.34^{+0.54}_{-0.54}$ & $-1.51^{+0.36}_{-0.36}$ & --                      & --\\ 
                SN 2013cs & 321 & $0.066^{+0.011}_{-0.011}$ & $6.05^{+0.10}_{-0.10}$ & $6.9^{+1.0}_{-0.9}$ & $7.66^{+0.43}_{-0.39}$ & $1.18^{+0.18}_{-0.19}$  & $7.67^{+0.78}_{-0.85}$ & $0.90^{+0.37}_{-0.40}$  & $10.42^{+0.93}_{-0.82}$ & $0.88^{+0.40}_{-0.37}$ \\
                SN 2012cg & 357 & $0.051^{+0.005}_{-0.005}$ & $5.81^{+0.12}_{-0.10}$ & $6.5^{+0.9}_{-0.9}$ & $7.68^{+0.24}_{-0.22}$ & $-1.48^{+0.10}_{-0.10}$ & $7.36^{+0.89}_{-0.91}$ & $-0.84^{+0.35}_{-0.39}$ & $12.30^{+1.38}_{-1.21}$ & $0.60^{+0.57}_{-0.58}$\\
                SN 1998bu & 364 & $0.047^{+0.011}_{-0.011}$ & $5.63^{+0.19}_{-0.12}$ & $5.5^{+0.7}_{-0.8}$ & $9.08^{+0.44}_{-0.40}$ & $-1.08^{+0.16}_{-0.16}$ & $8.52^{+0.54}_{-0.48}$ & $-1.31^{+0.97}_{-0.88}$ & $12.41^{+1.37}_{-1.21}$ & $-1.03^{+0.68}_{-0.59}$  \\ 
                SN 2012fr & 375 & $0.038^{+0.004}_{-0.004}$ & $5.77^{+0.16}_{-0.19}$ & $5.7^{+1.5}_{-0.8}$ & $7.17^{+0.19}_{-0.18}$ & $1.42^{+0.10}_{-0.10}$  & $6.97^{+0.97}_{-0.63}$ & $0.65^{+0.28}_{-0.30}$  & $10.16^{+1.12}_{-0.95}$ & $0.58^{+0.55}_{-0.59}$  \\
                SN 2013aa & 378 & $0.035^{+0.003}_{-0.003}$ & $5.67^{+0.14}_{-0.11}$ & $6.6^{+1.1}_{-1.1}$ & $7.32^{+0.24}_{-0.22}$ & $-1.35^{+0.10}_{-0.09}$ & $7.51^{+0.71}_{-0.65}$ & $-0.91^{+028}_{-0.29}$  & $10.29^{+1.57}_{-1.08}$ & $-0.13^{+0.53}_{-0.55}$\\
                SN 2013aa & 443 & $0.025^{+0.003}_{-0.003}$ & $5.44^{+0.28}_{-0.16}$ & $5.0^{+0.7}_{-0.9}$ & $7.87^{+0.18}_{-0.17}$ & $-1.18^{+0.07}_{-0.07}$ & $8.19^{+0.46}_{-0.48}$ & $-1.06^{+0.22}_{-0.22}$ & $12.07^{+1.86}_{-1.88}$ & $0.15^{+0.92}_{-0.86}$\\
                SN 2014J  & 468 & $0.020^{+0.002}_{-0.002}$ & $5.24^{+0.28}_{-0.16}$ & $4.3^{+0.7}_{-0.9}$ & $8.59^{+0.14}_{-0.14}$ & $0.45^{+0.06}_{-0.06}$  & $9.13^{+0.31}_{-0.32}$ & $0.36^{+0.13}_{-0.13}$  & --                      & --\\ 
                SN 2014J  & 496 & $0.015^{+0.002}_{-0.002}$ & $5.15^{+0.14}_{-0.15}$ & $3.8^{+0.4}_{-0.3}$ & $9.15^{+0.14}_{-0.14}$ & $0.67^{+0.07}_{-0.07}$  & $8.75^{+0.77}_{-0.66}$ & $0.03^{+0.27}_{-0.29}$  & --                      & --\\ 
                \\
                \hline
                Mean + Std\tablefootmark{d} & & & & & $7.9\pm0.8$ & $1.1\pm0.5$ & $7.9\pm1.0$ & $0.9\pm0.7$ & $10.7\pm1.8$ & $0.5\pm0.6$\\
                \hline
        \end{tabular} 
        }
        \tablefoot{
        For SN\,1998bu at 269 days the temperature and density are not fitted (see text for details).\\
        \tablefoottext{a}{Days after the explosion, assuming a rise time of $\sim 18$ days \citep{2011MNRAS.416.2607G}.}
        \tablefoottext{b}{Doppler full width at half maximum.}
        \tablefoottext{c}{Shift velocity.}
        \tablefoottext{d}{Mean and standard deviation over all fit samples. For shift velocities we calculated the mean and   standard deviation of the absolute values.}
        }
\end{table*}

\subsection{Shifts and widths of singly and doubly ionised material}
\label{sectionShiftsWidths}
\begin{figure}
        \centering
        \resizebox{\hsize}{!}{\includegraphics{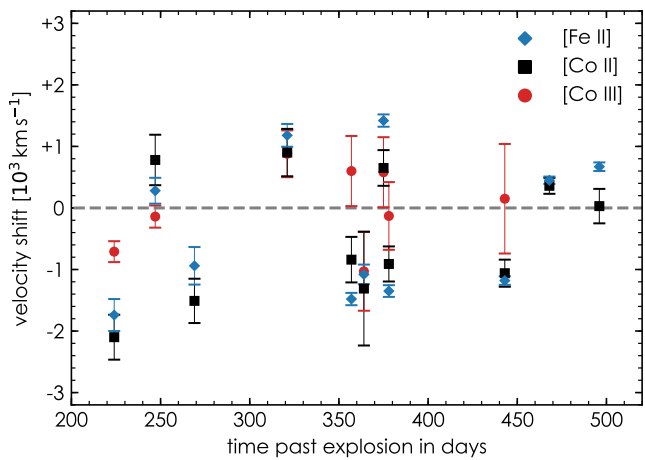}}
        \caption{Velocity shifts of [\ion{Fe}{II}], [\ion{Co}{II}], and [\ion{Co}{III}] as a function of time after explosion.} 
        \label{Shifts}
\end{figure}

\begin{figure}
        \centering
        \resizebox{\hsize}{!}{\includegraphics{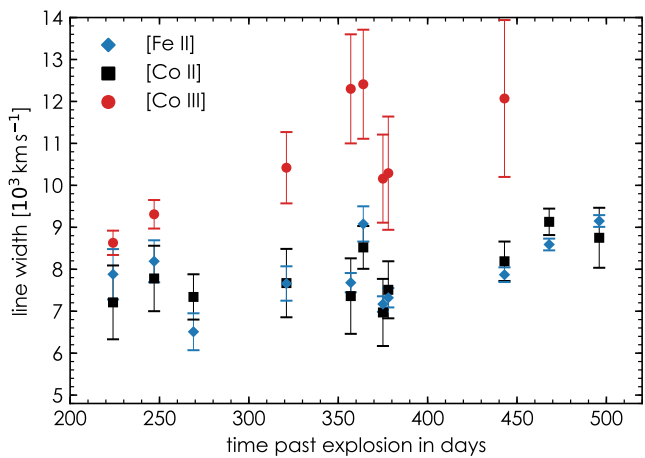}}
        \caption{Line widths of [\ion{Fe}{II}], [\ion{Co}{II}], and [\ion{Co}{III}] as a function of time after explosion.} 
        \label{Widths}
\end{figure}

We can use the velocity shifts of [\ion{Fe}{II}], [\ion{Co}{II}], and [\ion{Co}{III}] to perform a similar study to \citet{2018MNRAS.477.3567M}. We present our results in Figs.~\ref{Shifts} and~\ref{Widths}. Different velocity shifts between stable material ([\ion{Ni}{II}]), singly ionised decay products ([\ion{Fe}{II}], [\ion{Co}{II}]) and doubly ionised decay products ([\ion{Fe}{III}], [\ion{Co}{III}]) would suggest their production in spatially separated regions of the ejecta \citep{2010ApJ...708.1703M}. As we fit the forbidden \ion{Co}{II} lines at $1.019\,\mu$m and $1.547\,\mu$m we can directly compare velocity offsets and widths of singly and doubly ionised cobalt. For \ion{Fe}{III} and \ion{Fe}{II} this cannot be done as the $4700$\,\AA\,\ion{Fe}{III}-dominated feature exhibits a shift of its central wavelength with time.

We find that the absolute velocity shifts of [\ion{Fe}{II}], $(1.1\,\pm\,0.5)\times10^3$\,km\,s$^{-1}$, are very similar to the absolute velocity shifts of [\ion{Co}{II}], $(0.9\,\pm\,0.7)\times10^3$\,km\,s$^{-1}$. Deviations of the velocity shift between the two ions are less than $500$\,km\,s$^{-1}$ for most spectra and likely the result of noise near the $1.019\,\mu$m [\ion{Co}{II}] feature. Velocity shifts of singly ionised ions can be either positive or negative for the observed sample. 

Lines of \ion{Fe}{II} and \ion{Co}{II} exhibit comparable widths. Line widths of [\ion{Co}{II}] usually have a higher uncertainty than line widths of [\ion{Fe}{II}] as there is only one unblended \ion{Co}{II} feature in our spectra, but multiple features of [\ion{Fe}{II}] can be found at $7\,200\,$\AA, $8\,600\,$\AA, $12\,600\,$\AA\,and $16\,400\,$\AA. We find widths of $(7.9\,\pm\,1.0)\times10^3$\,km\,s$^{-1}$ for [\ion{Co}{II}] and $(7.9\,\pm\,0.8)\times10^3$\,km\,s$^{-1}$ for [\ion{Fe}{II}].

We find that [\ion{Co}{III}] shows lower velocity offsets from its rest wavelength than singly ionised iron and cobalt. While there are spectra in which the [\ion{Co}{III}] velocity shift is similar to [\ion{Fe}{II}] and [\ion{Co}{II}] (SN\,2013cs, SN\,1998bu), for the majority of spectra in our sample we get different velocity shifts between [\ion{Co}{III}] and [\ion{Co}{II}]. We find $(0.5\,\pm\,0.6)\times10^3$\,km\,s$^{-1}$  for the mean absolute  [\ion{Co}{III}] velocity shift, compared to $(0.9\,\pm\,0.7)\times10^3$\,km\,s$^{-1}$ for [\ion{Co}{II}]. As noted in \citet{2018MNRAS.477.3567M}, shifts of [\ion{Co}{III}] seem to be consistent with zero offset from the rest wavelength. 

On average, [\ion{Co}{III}] lines also appear to be broader than [\ion{Co}{II}] lines by about $3\,000$\,km\,s$^{-1}$ ($(10.7\,\pm\,1.8)\times10^3$\,km\,s$^{-1}$ for [\ion{Co}{III}] compared to $(7.9\,\pm\,1.0)\times10^3$\,km\,s$^{-1}$ for [\ion{Co}{II}]). We note that supernovae which exhibit a strong continuum in the $5\,900\,$\AA\,region (SN\,2012cg, SN\,1998bu at 364\,d, SN\,2013aa at 443\,d) tend to have higher line widths than supernovae without continuum in this region. In these cases, the line widths appear to be strongly affected by the manual continuum subtraction. [\ion{Co}{III}] widths determined in this work are compatible within their uncertainties with results from \citet{2018MNRAS.477.3567M}. [\ion{Co}{III}] lines in our fits are redshifted by $\sim 500\,$km\,s$^{-1}$ compared to \citet{2018MNRAS.477.3567M}. This systematic shift is mostly due to the presence of [\ion{Fe}{II}] lines from the a$^4$F$\,-\,$a$^2$D multiplet ($5335, 5377, 5528$\,\AA) on the blue side of the $5\,900\,$\AA\,feature which are not included in the work of \citet{2018MNRAS.477.3567M}. We note that the choice of the continuum differs between this work and \citet{2018MNRAS.477.3567M}.

\section{Comparison of  M$_{\ion{Co}{II}}$\,/\,M$_{\ion{Fe}{II}}$ with explosion model yields}
\label{SectionBayes}

Besides \element[][56]{Ni}, explosion models predict the presence of the isotopes \element[][54,56]{Fe}, \element[][57]{Ni} and \element[][58]{Ni}. If the explosive burning in the white dwarf occurs at high densities (e.g. M$_{Ch}$ models) stable \element[][54,56]{Fe} and \element[][58]{Ni} and rapidly decaying \element[][57]{Ni} can be produced directly. Sub-M$_{Ch}$ models lack a high enough central density to produce neutron-rich material. However, if the progenitor has a high metallicity (several times solar) there is an excess of neutrons, resulting in the synthesis of \element[][54,56]{Fe}, \element[][57]{Ni,} and stable \element[][58]{Ni} with masses comparable to those of M$_{Ch}$ explosions \citep{2010ApJ...714L..52S,2013MNRAS.429.1425R}. 

In Fig.~\ref{decay} the prediction of the M$_{\ion{Co}{}}$\,/\,M$_{\ion{Fe}{}}$ ratio is shown in the case of pure \element[][56]{Ni} to \element[][56]{Co} to \element[][56]{Fe} decay and for nucleosynthetic yields of the M$_{Ch}$ deflagration model `W7 Z$_\odot$' \citep{2018SSRv..214...67N}, the M$_{Ch}$ delayed-detonation model `N100' \citep{2013MNRAS.429.1156S}, and the sub-M$_{Ch}$ CO detonation model `det\_2010\_1.06' \citep{2010ApJ...714L..52S}. The `W7 Z$_\odot$' model produces $2.8\,\%$ \element[][57]{Ni} and $35\,\%$ \element[][54,56]{Fe}, the M$_{Ch}$ delayed detonation model `N100' produces $3.1\,\%$ \element[][57]{Ni} and $19.4\,\%$ \element[][54,56]{Fe}, and the sub-M$_{Ch}$ CO detonation model `det\_2010\_1.06' produces $0.6\,\%$ \element[][57]{Ni} and no stable iron, each given in fractions of the \element[][56]{Ni} mass. As mentioned before, a high metallicity progenitor like in the `det\_2010\_1.06\_0.075Ne' model \citep{2010ApJ...714L..52S} produces similar masses of \element[][54,56]{Fe} and \element[][57]{Ni} as the `N100' model.

Stable iron produced in the explosion has the strongest effect on M$_{\text{Co}}$\,/\,M$_{\text{Fe}}$ while not all of the \element[][56]{Co} has yet decayed to \element[][56]{Fe}  ($<300\,$d). \element[][57]{Ni} decays within a few days to \element[][57]{Co}, which has a much longer half-life than \element[][56]{Co}:
\begin{align}
        &_{28}^{57}\text{Ni} \xrightarrow{\,\,\,\,T_{1/2}=1.48\,\text{d}\,\,\,\,}\, _{27}^{57}\text{Co}\xrightarrow{\,\,\,\,T_{1/2}=271.74\,\text{d}\,\,\,\,}\, _{26}^{57}\text{Fe.}\label{Ni57Decay}
\end{align}
Due to the longer half-life it can still be found late in the nebular phase ($>  \SI{400}{\day}$) when almost no \element[][56]{Co} remains.

\ion{Co}{II} and \ion{Fe}{II} have similar ionisation potentials and appear to be co-located within the ejecta (see Section \ref{SectionFitResults}). We therefore assume that the relative ionisation fraction between \ion{Co}{II} and \ion{Fe}{II} does not change with time, allowing us to generalise M$_{\ion{Co}{II}}\,/\,$M$_{\ion{Fe}{II}}$ to M$_{\ion{Co}{}}\,/\,$M$_{\ion{Fe}{}}$. This ratio measures the total mass of cobalt to iron at a given time, with possible contributions from stable \element[][54,56]{Fe} and decay products according to Eqs.~\ref{Ni56Decay} and~\ref{Ni57Decay}. We compare the predicted M$_{\ion{Co}{}}\,/\,$M$_{\ion{Fe}{}}$ ratio of different models with the results from our fits by determining their likelihood based on their Bayes factors. The Bayes factor quantifies the evidence of data $D$ for model $M_a$ compared to model $M_b$. In our analysis we consider the nested models $M_0$ to $M_3$ (see Table~\ref{tableModels}). The possible parameters of these models are initial abundances of the isotopes \element[][56]{Ni}, \element[][54,56]{Fe,} and \element[][57]{Ni}.

\begin{table} 
        \caption{Models and their included isotopes produced in the explosion.} 
        \label{tableModels}     
        \centering 
        \footnotesize
        \begin{tabular}{c c  c c} 
        \hline
        \hline 
        Model & \element[][56]{Ni} &  $\element[][54,56]{Fe}$ & $\element[][57]{Ni}$ \\      
                \hline  
                $M_0$ & \cmark             & \xmark                   & \xmark               \\
                $M_1$ & \cmark             & \cmark                   & \xmark               \\
                $M_2$ & \cmark             & \xmark                   & \cmark               \\
                $M_3$ & \cmark             & \cmark                   & \cmark               \\
                \hline
        \end{tabular}
\end{table}

\begin{figure}
        \centering
        \resizebox{\hsize}{!}{\includegraphics{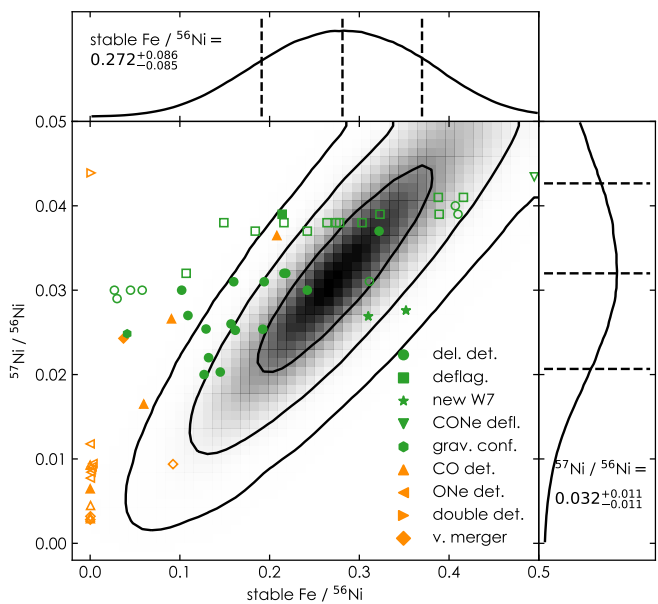}}
        \caption{
        Covariances and normalized posterior densities of model $M_3$. Theoretical predictions of the initial \element[][54,56]{Fe} and \element[][57]{Ni} fraction of various explosion models are indicated by green (Near-Ch-mass) and orange (sub-Ch-mass) symbols. Theoretical predictions with too little ($<0.4\,$M$_\odot$) or too much ($>0.8\,$M$_\odot$) \element[][56]{Ni} compared to Branch-normal SNe Ia are marked as empty symbols. Black lines show the 1, 2, and 3$\sigma$ credibility regions. Numbers next to the 1D histograms indicate the $68$\,\% credibility range of the individual parameters. \newline
        Models: Del. Det. \citep{2013MNRAS.429.1156S,2014A&A...572A..57O,2018SSRv..214...67N}, Deflag. \citep{2014MNRAS.438.1762F}, new W7 \citep{2018SSRv..214...67N}, CONe Defl. \citep{2015MNRAS.450.3045K}, Grav. Conf. \citep{2016A&A...592A..57S}, CO Det. \citep{2010ApJ...714L..52S,2018SSRv..214...67N}, ONe Det. \citep{2015A&A...580A.118M}, Double Det. \citep{2012MNRAS.420.3003S}, V. Merger \citep{2010Natur.463...61P, 2012ApJ...747L..10P,2013ApJ...778L..18K,2016MNRAS.459.4428K}
    }
        \label{bayesplots}
\end{figure}

\begin{figure}
        \centering
        \resizebox{\hsize}{!}{\includegraphics{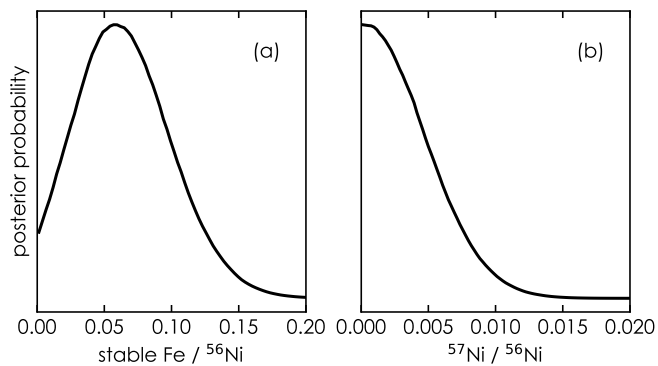}}
        \caption{
        (a) Normalized posterior density of model $M_1$ with only stable iron \element[][54,56]{Fe} and radioactive \element[][56]{Ni}. (b) Normalized posterior density of model $M_2$ with only  radioactive \element[][57]{Ni} and \element[][56]{Ni}.
        }
        \label{bayesplots2}
\end{figure}

For nested models, the Bayes factor can be computed analytically if the priors are separable and the more complicated model $M_b$ becomes the simpler model $M_a$ for $\theta = \theta_0$:
\begin{equation}
        B_{ab} = \frac{P(\theta = \theta_0\vert D, M_b)}{P(\theta = \theta_0\vert M_b)}
.\end{equation}
We use the mass ratio results from our individual spectral fits to compute the Bayes factor between models $M_0$ and $M_3$. Prior odds are set to one as we do not have an initial preference for any model. For \element[][54,56]{Fe} and \element[][57]{Ni} we adopt flat priors $0 \leq \text{M}_{\element[][54,56]{Fe}}\,/\,\text{M}_{\element[][56]{Ni}} \leq 0.5$ and $0 \leq \text{M}_{\element[][57]{Ni}}\,/\,\text{M}_{\element[][56]{Ni}} \leq 0.05$. The covariances and posterior densities of the models are shown in Fig.~\ref{bayesplots} for $M_3$ and in Fig.~\ref{bayesplots2} for $M_1$ and $M_2$. Due to the introduction of additional degrees of freedom all models fit the data reasonably well.

\begin{table} 
        \caption{Bayes factors and their interpretation between the nested models and relative probabilities of the models.} 
        \tiny
        \label{tableBayesFactor}        
        \centering 
    \scalebox{.95}{
        \begin{tabular}{ll}
                \begin{tabular}{c c c c} 
                        \hline
                        \hline 
                        Bayes F.\tablefootmark{a} &  ln(B) & Pref.\tablefootmark{b} & Interpr.\tablefootmark{c} \\  
                        \hline  
                        $B_{01}$                  & $+0.6$ & $M_0$                  & weak                      \\
                        $B_{02}$                  & $+2.1$ & $M_0$                  & positive                      \\
                        $B_{03}$                  & $-3.8$ & $M_3$                  & strong                  \\
                        $B_{13}$                  & $-4.1$ & $M_3$                  & strong                    \\
                        $B_{23}$                  & $-5.8$ & $M_3$                  & very strong                    \\
                        \hline
                \end{tabular}
                &
                \begin{tabular}{c c} 
                        \hline
                \hline 
                        Model   & Rel. Prob.\tablefootmark{d} \\        
                        \hline  
                        $M_{0}$ & $0.022$                                         \\
                        $M_{1}$ & $0.013$                                         \\
                        $M_{2}$ & $0.002$                                         \\
                        $M_{3}$ & $0.963$                                         \\
                                &                                                         \\
                        \hline
                \end{tabular}
        \end{tabular}
    }
        \tablefoot{
                \tablefoottext{a}{Bayes Factor $B_{ab}$ between models $M_a$ and $M_b$.}
                \tablefoottext{b}{Preferred model.}
                \tablefoottext{c}{Interpretation according to \citet{doi:10.1080/01621459.1995.10476572}.}
                \tablefoottext{d}{Relative probability $P(D\vert M_a)$\,/\,$\sum P(D\vert M_i)$.}
        }
\end{table}
To determine the best fitting model given the data we compute the Bayes factors between the models as shown in Table \ref{tableBayesFactor}. The M$_{\ion{Co}{}}\,/\,$M$_{\ion{Fe}{}}$ ratio of both M$_{Ch}$ and sub-M$_{Ch}$ explosion models follows closely the mass ratio from pure \element[][56]{Ni} decay between 250 and 350 days after the explosion. Data points between these epochs cannot distinguish between either model.

Most M$_{Ch}$ explosion models that predict between $0.4$ and $0.8\,$M$_\odot$ of \element[][56]{Ni} (filled symbols in Fig.~\ref{bayesplots}) lie in the $3\sigma$ error ellipse of model $M_3$. The error ellipse also contains the aforementioned high-metallicity CO detonation model `det\_2010\_1.06\_0.075Ne' \citep{2010ApJ...714L..52S}. Models, that yield less than 0.4\,M$_{\odot}$ or more than 0.8\,M$_{\odot}$ of \element[][56]{Ni} are marked with empty symbols. The models with only \element[][57]{Ni} or only \element[][54,56]{Fe} in addition to the \element[][56]{Ni} are not allowed as the gain in fit quality is small compared to the added parameter space (see Fig.~\ref{bayesplots2}). A comparison between models \element[][56]{Ni} only ($M_{0}$) and our more complex $M_{3}$ model favours the complex model.

\section{Discussion}
\label{SectionDiscussion}
\subsection{Emission line widths and velocity shifts}

\citet{2018MNRAS.477.3567M} found that the velocity shifts and line widths of [\ion{Fe}{II}] and [\ion{Ni}{II}] of SNe in the sample are compatible within their uncertainties. We measure an average [\ion{Fe}{II}] velocity shift of $(1.1\,\pm\,0.5)\times10^3$\,km\,s$^{-1}$, across the sample, compared to $(0.9\,\pm\,0.7)\times10^3$\,km\,s$^{-1}$ for [\ion{Co}{II}]. Lines of [\ion{Fe}{II}] and [\ion{Co}{II}] are either both redshifted or both blueshifted. The average line widths are $(7.9\,\pm\,1.0)\times10^3$\,km\,s$^{-1}$ in the case of [\ion{Co}{II}] and $(7.9\,\pm\,0.8)\times10^3$\,km\,s$^{-1}$ for [\ion{Fe}{II}]. This indicates that the emission from all singly ionised iron-group ions originates from the same spatial region in the ejecta. The emitting region of singly ionised material appears to be located off-centre and is smaller than the emitting region of doubly ionised cobalt by about $3\,000\,$km\,s$^{-1}$. We do not find significant velocity shift and line width offsets between [\ion{Fe}{II}] emission in the optical and NIR. We confirm the results of \citet{2018MNRAS.477.3567M} that the absolute velocity shift of [\ion{Co}{III}] is lower than that of the singly ionised species and for most spectra in our sample compatible with no offset from the rest wavelength. 

Nucleosynthetic calculations for delayed detonations including multiple ignition points by \citet{2013MNRAS.429.1156S} show the stable iron group \element[][54]{Fe}, \element[][56]{Fe,} and \element[][58]{Ni} to be distributed within the supernova in close association with \element[][56]{Ni}. At the epochs of our observations ($\sim 200$ to 500 days), the $\gamma$-rays still deposit energy into the ejecta and any stable iron, whether in the core or distributed within the ejecta, should be heated and contribute to the emission. If stable and radioactive materials are co-located in the same region, the stable material can also be heated by positrons. In fact, evidence for heating of stable elements is found in the late-time optical spectra where the line of [\ion{Ni}{II}] is evident at 7\,378\,\AA\,\citep[e.g.][]{2010ApJ...708.1703M,2018MNRAS.477.3567M}. No significant amount of \element[][56]{Ni} can be expected to be present in the ejecta 100 days after the explosion. 

It is possible to hide a large amount of material if it is not heated and therefore not emitting. At epochs later than 350 days, when the mean free path for the $\gamma$-rays from the decay of \element[][56]{Co} is larger than the radius of the supernova and the bulk of the $\gamma$-rays do not deposit their energy in the ejecta, the energy deposition within the ejecta is almost exclusively via the $\beta^+$ channel \citep[e.g.][]{1980PhDT.........1A,1989ApJ...346..395W,1992ApJ...401...49L,2007ApJ...662..487W}. It is assumed, due to trapping by magnetic fields, that the positron energy deposition is local to the emission region. It is, therefore, possible to have pockets of cold gas. In such a case those regions would not contribute to the spectra. Unlike the observations of SN\,2003hv \citep{2006ApJ...652L.101M,2010ApJ...708.1703M,2011MNRAS.416..881M}, as discussed in \citet{2018MNRAS.477.3567M} our spectra do not exhibit flat-topped profiles of emission features and there is no evidence for an ashes bubble in the centre of, or off-centre, in the ejecta. A more sophisticated analysis of the line profiles of these species in the context of energy deposition scenarios and distributions can be found in \citet{2018ApJ...861..119D}. 

\subsection{The origin of iron and cobalt}

Doubly ionised cobalt and iron lines have been used in the past by other authors \citep[see][]{1994ApJ...426L..89K} to derive that the mass ratio M$_{\ion{Co}{}}\,/\,$M$_{\ion{Fe}{}}$ in SNe Ia is governed by the radioactive decay of \element[][56]{Co} to \element[][56]{Fe}. That work relied on iron and cobalt lines below $5\,500\,$\AA. Singly ionised lines above $7\,000\,$\AA\,become optically thin after about $150$ days. 

Our work determines that the singly ionised lines observed in the spectra arise from the daughter products of the radioactive \element[][56]{Ni}. Even though we can only constrain the temperature weakly, it is evident that the mass ratio evolution is not a temperature effect. While it is plausible to assume variations in the temperature of the ejecta from one SN to another, these are unlikely to be extreme as the heating and cooling are fundamentally dictated by the same elements (the daughter products of the decay of \element[][56]{Ni}). Furthermore, the production of \element[][56]{Ni} is tightly linked to the density and temperature of the progenitor. Simply put, the inner ejecta of normal SNe Ia should be very similar in terms of their excitation within the boundaries of this analysis. This is borne out by the very strong similarity of the emission line spectra obtained during the first year after the explosion \citep{1998ApJ...499L..49M}.

By fitting the spectra with forbidden emission lines of our set of NLTE atoms we find a decline of the temperature in the ejecta from $\sim8\,000$\,K at 200 days to $\sim4\,000$\,K at 500 days. Such a temperature evolution is consistent with predictions from simulations \citep{2015ApJ...814L...2F}. We also find that the density of the emitting material decreases from a few times $10^6$\,cm$^{-3}$ to about $10^5$\,cm$^{-3}$ between the youngest and oldest spectrum in our sample. These densities are consistent with a simple distribution of roughly half a solar mass of singly and doubly ionised iron group material in a volume expanding at $\sim 8\,000$\,km\,s$^{-1}$ for the age of the SN. 

We ruled out models containing only stable iron or only \element[][57]{Ni} in addition to \element[][56]{Ni}. Theoretical predictions of the cobalt to iron mass ratio from M$_{Ch}$ and sub-M$_{Ch}$ explosion simulations differ only slightly from the pure \element[][56]{Ni} decay chain. Within the Bayesian framework we find evidence ($B_{03}=-3.8$) for a model with \element[][56]{Ni}, stable iron and \element[][57]{Ni}  produced during the explosion similar to most M$_{Ch}$ models, compared to a purely \element[][56]{Ni} origin of iron. 

Due to largely unknown ejecta densities and temperatures, the uncertainties of the derived cobalt to iron mass ratio are quite large. Future observations of the relative strength of the [\ion{Co}{II}] 10.52\,$\mu$m, 14.74\,$\mu$m and 15.46\,$\mu$m lines (e.g. with $JWST$) would also allow for a direct measurement of the density and temperature of the emitting material in the range $10^4$ -- $10^6$\,cm$^{-3}$ and 3\,000 -- 10\,000\,K \citep{2016MNRAS.456.1974S}. A more accurate \ion{Co}{II}\,/\,\ion{Fe}{II} mass ratio could be used to determine the amount of \element[][57]{Ni} and \element[][54,56]{Fe} for individual SNe rather than assuming similar progenitor scenarios for a sample of SNe.

The presence of \element[][57]{Ni} was also discussed for the nearby SN\,2011fe \citep{2017MNRAS.468.3798D,2017MNRAS.472.2534K,2017ApJ...841...48S} and SN\,2012cg \citep{2016ApJ...819...31G}. For these two objects photometric measurements were obtained at extremely late phases ($>1\,000\,$days after explosion) to construct a pseudo-bolometric light curve. \citet{2016ApJ...819...31G} claimed that \element[][57]{Ni} is required to explain the late-time light curve of SN\,2012cg. Both \citet{2017MNRAS.468.3798D} and \citet{2017ApJ...841...48S} argued for a detection of \element[][57]{Ni} in SN\,2011fe, albeit with very different abundances, pointing to either near-M$_{Ch}$ \citep{2017MNRAS.468.3798D} or sub-M$_{Ch}$ explosions \citep{2017ApJ...841...48S}. \citet{2017MNRAS.472.2534K} tested the effect of various isotopic abundances and physical processes on the light curve and came to the conclusion that one cannot determine the \element[][57]{Ni} abundance from the light curve without detecting the mid-IR cooling lines.

\section{Conclusions}

We fitted late time spectra of SNe Ia with broadened emission profiles from NLTE level populations using a Bayesian sampler. We computed good fits with similar widths and velocity offsets among the singly ionised atoms, indicating a common emission region. We find that doubly ionised cobalt is located in a broader and more centred region of the ejecta. We have shown that the singly ionised iron group elements in the late-time spectra of SNe Ia change their flux ratios in agreement with an evolution of their mass ratios that would be governed by the radioactive decay of \element[][56]{Co} to \element[][56]{Fe}. This result disfavours the presence of only stable iron or only the decay products of \element[][57]{Ni} in the ejecta in addition to \element[][56]{Ni}, at least for the SNe observed here. For a model that produced both stable iron and \element[][57]{Ni} in addition to \element[][56]{Ni} during the explosion we find $^{54,56}$Fe\,/\,$^{56}$Ni $=0.272\pm0.086$ and $^{57}$Ni\,/\,$^{56}$Ni $=0.032\pm0.011$ with a ratio of stable iron to \element[][57]{Ni} of $\sim 8.5:1$. 

\begin{acknowledgements}
We would like to thank the anonymous referee for the helpful comments.
We thank the staff at Gemini and Paranal observatories. This research would not be possible without their efforts in supporting queue/service mode observing. This research has made use of the NASA/IPAC Extragalactic Database (NED) which is operated by the Jet Propulsion Laboratory, California Institute of Technology, under contract with the National Aeronautics and Space Administration. This work made use of the Heidelberg Supernova Model Archive (HESMA), https://hesma.h-its.org. We would like to thank Johannes Buchner for helpful discussions and comments. AF acknowledges the support of an ESO Studentship. KM is supported by the UK STFC through an Ernest Rutherford fellowship. ST acknowledges support by TRR33 'The Dark Universe' of the German Research Foundation (DFG). WEK acknowledges the Excellence Cluster Universe, Technische Universit\"at M\"unchen, Boltzmannstrasse 2, D-85748 Garching, Germany and acknowledges the support of an ESO Fellowship.

\end{acknowledgements}
\bibliographystyle{aa} 
\bibliography{bib} 
\end{document}